\documentclass[prd,onecolumn,superscriptaddress,floatfix,longbibliography]{revtex4}%nofootinbib,floatfix

\usepackage{amsfonts}
\usepackage{amsmath}
\usepackage{graphicx}
\usepackage{amssymb}
\usepackage{color}
\usepackage{subfigure}
\usepackage{mathtools}

\newcommand{\Melement}[3]{\left\langle #1 \right| #3 \left| #2 \right\rangle } % for Matrix Elements

\begin{document}
\title{Noise and dissipation on a moving mirror induced by the dynamical Casimir emission}
%%Master equation for a moving mirror microscopically coupled to a field
%%Open quantum system approach for the back-raction onto a movign mirror by DCE
%
\author{Salvatore Butera}
\affiliation{School of Physics and Astronomy, University of Glasgow, Glasgow G12 8QQ, United Kingdom}

\begin{abstract}
We adopt an open quantum system approach to study the effects of the back-reaction from a quantum field onto the dynamics of a moving mirror. We describe the coupling between the mirror and the field by using a microscopic model from which the dielectric response of the mirror is obtained from first principles. Using second-order perturbation theory, we derive the master equation governing the mechanical motion of the mirror. Our analysis reveals that the mirror experiences coloured noise and non-local dissipation, which originate from the emission of particle pairs via the dynamical Casimir effect. We show that the noise and dissipation kernels, that enter in the definition of the time-dependent coefficients of the master equation, are related by fluctuation-dissipation relations.
\end{abstract}
\maketitle

\section{Introduction}
Quantum optomechanics is the field of research that studies the interplay between mechanical motion and light, at the quantum level~\cite{Aspelmeyer_RMP}. The field has flourished over the past few decades, since progress in nano-fabrication technologies and cooling techniques has made possible reaching the regime in which quantum fluctuations represent the predominant source of noise and have sizeable effects on the dynamics of mechanical resonators~\cite{Cleland2010,chan2011laser,Aspelmeyer-Science-2020}. Optomechanical systems are nowadays among the most promising platforms for the development of future quantum technologies~\cite{Paternostro_Review}. They can be used for example in metrology, as ultra-high-precision sensors and actuators of nano-metric motion~\cite{stange2021science}, and have been recently proposed as a platform to implement quantum thermal machines~\cite{Nori-QuantumHEatEng-2007,Bariani-QuantumHEatEng-I,Bariani-QuantumHEatEng-II,Bariani-QuantumHEatEng-III,Nori-QuantumHEatEng-2023}.
Remarkably, thanks to their exceptionally long coherence time~\cite{MacCabe2020,Beccari2022}, quantum-enabled acoustic resonators represent a promising alternative to superconducting circuits for quantum information storage~\cite{Pistolesi-PRX-2021,navarathna2022good} and processing~\cite{Rabl2012}. It is clear that underpinning the use of these systems in quantum technology is the ability to exploit the coupling with optical or microwave radiation field to transfer information in and out of the acoustic resonators, thus implementing coherent control of mechanical vibrations at the level of few or even single quantum excitations~\cite{lahaye2009nanomechanical,von2022parity}. It is thus crucial for the research community to develop accurate theoretical models that describe the dynamics of nano-scale devices and their interaction with the radiation field.

At a more fundamental level, quantum fields interacting with movable boundary conditions provide a convenient framework to investigate the physics of the quantum vacuum and its instability in certain physical conditions. As originally demonstrated by Moore~\cite{Moore-DCE-1970}, Fulling and Davies~\cite{Fulling_Davies-DCE,Davies-Fulling-BH}, the vacuum fluctuations of a quantum field can be converted into pairs of real photons by the non-adiabatic motion of a boundary condition (mimicking an ideal, perfectly reflecting mirror). This effect is known in the literature as dynamical Casimir effect (DCE)~\cite{Yablonovitch-DCE,Schwinger-DCE} and has been extensively investigated in the literature~\cite{dodonov2020-DCE}. The physical mechanism that engenders the dynamical Casimir emission is equivalent to the process by which particles are created in an expanding universe, as anticipated by Parker~\cite{Parker-PartCr-I,Parker-PartCr-II}, and is related to the evaporation of black holes by the emission of Hawking radiation~\cite{Hawking1974,Hawking1975}.
In spite of the impressive recent technological advances in optomechanics, experimental evidence of the excitation of the electromagnetic vacuum, triggered by mechanical motion, has so far evaded observation. Experimental evidence of the DCE has been achieved instead in the context of analogue models based on superconducting circuit-QED~\cite{Wilson-DCE-Analog-2011,Lahteenmaki-DCE-Analog-2013}, in which case the mechanical motion of the mirror is mimicked by modulating in time the effective electric length of a microwave cavity, at frequencies high enough to resonantly amplify the DCE signal to a sizeable amplitude~\cite{Johansson-PRL-2009,Johansson-PRA-2010}.

According to its original formulation, the DCE is a test-field effect, meaning that the time-dependent boundary condition imposed to the field is not a dynamical degree-of-freedom of the system, but is a mere external parameter. In a real optomechanical system instead, the mechanical objects that interact with radiation are engendered with their own quantum dynamics and the effects of the back-action (back-reaction) of the quantum field onto the mechanical motion need to be taken in to account~\cite{kippenberg2008cavity}. In static configurations, early works \cite{Giulio-PRL-2013,Armata-PRD-2015,Bartolo2015,Armata-PRD-2017} have studied the modification of the ground state of an optomechanical system, due to the quantized motion of the optical mirror, while the role of the quantum vibration of an acoustic resonator as mediator of the interaction between two physically separated fields have been recently investigated in Ref.~\cite{Montalbano-2023}.
Further studies investigated out-of-equilibrium configurations, demonstrating that the electromagnetic vacuum is able to affect the motion of mechanical objects~\cite{KardarRMP1999}, inducing dissipation and fluctuations~\cite{Reynaud-QuantOpt-1992,Reynaud-JdP-1993,Reynaud-PhysLettA-1993,Unruh-PRD-2014,Savasta-PRX-2018,Butera-PRA-2019}, as well as quantum decoherence~\cite{Dalvit-PRL-2000,MaiaNeto-PRA-2000,Butera-EPL-2019}. Such decoherence effects are responsible for driving the quantum-to-classical transition of any macroscopic objects interacting with light~\cite{zurek1991decoherence}.
Pioneering theoretical works pushed these studies towards more complex configurations, proving vacuum mediated coherent exchange of excitations at the single phonons level between physically separated mirrors~\cite{Savasta-PRL-2019,Butera-InfFunc-2022} and between mirrors and atoms~\cite{Nori2023}, as well as photons hopping induced by mechanical zero-point fluctuations~\cite{Savasta-Hopping-2023} in optomechanical lattices~\cite{Clerk-OptoMechLatt-2014}. 

This paper adds to the existing literature, with the objective of setting up the formalism to address the back-reaction by DCE, adopting an open quantum system approach to the problem. This approach is analogous to the one developed in Refs.~\cite{Dalvit-PRL-2000,MaiaNeto-PRA-2000}, but makes use of the more refined microscopic model developed in Ref.~\cite{MOF1}, to describe the interaction between mirrors and light. By using this model, we are able to account for the dielectric properties of the mirror from first principles, achieving a description of the interaction beyond the standard theory of radiation pressure~\cite{Law-MirFieldInt-1995,Law-MirFieldInt-2011}, which relies on unphysical boundary conditions enforcing the field to vanish in correspondence of the mirrors. This naturally leads to an ultraviolet cut-off in the theory, thus solving the pathologic ultraviolet behaviour that plagues standard theory of radiation pressure in the case of the multi-mode field configurations~\cite{Giulio-PRL-2013,Armata-PRD-2015,Bartolo2015,Armata-PRD-2017,Montalbano-2023}. For simplicity, we develop a one-dimensional (1D) model, so that the electromagnetic vector potential can be treated as a scalar field. Despite this simplifying assumption, we expect that the theory we present is able to capture all the relevant qualitative features of the back-reaction. A similar study has recently been presented in Ref.~\cite{Kanu-PRD-2021}, where the authors attempted to use the theory of the influence functionals in order to find an effective action for the mechanical motion of a mirror interacting with the radiation field.

The paper is organized as follows: In Sec.~\ref{Sec:MOF}, we review the microscopic model proposed in Ref.~\cite{MOF1}, that we use to describe the mirror-field interaction. There, we consider the mirror being static, since the focus is on deriving the dielectric response of the mirror to radiation. In Sec.~\ref{Sec:MovingMirror}, we consider the case of the mirror vibrating within a confining harmonic potential, and present the Hamiltonian that describes its dynamics. In Sec.~\ref{Sec:ME}, we use perturbation theory to derive the master equation for the reduced density matrix of the mirror and introduce key physical quantities, such as the noise and dissipation kernels, that describe the back-reaction effects of the field onto the mechanical motion of the mirror, at the microscopic level. This is the first major result of this paper. At the macroscopic level, the back-reaction appears in the form of an effective friction experienced by the mirror, as well as diffusion, the latter being responsible for quantum decoherence. Fluctuation and dissipation kernels are defined in terms of relevant physical properties of the environment in which the mirror is located (i.e, the field and the microscopic degree-of-freedom that mediates the interaction), that are encoded in the correlation functions calculated in Sec.~\ref{Sec:NoiseDiss}. Specifically, in Secs.~\ref{Sec:Corr+} and \ref{Sec:Corr-q}, we determine the correlation functions for the field and the internal degree-of-freedom of the mirror, while in Sec.~\ref{Sec:NuMuM} we present the explicit expressions for the noise and dissipation kernels for the mechanical motion of the mirror, which are the second key result of this paper. Furthermore, we give evidence that the back-reaction onto the moving mirror is associated with the emission of pairs of particles by DCE. In Sec.~\ref{Sec:Reps}, we present the master equation both in the position and Wigner representation, and discuss in more details the physical processes it accounts for. Our conclusions are finally drawn in Sec.~\ref{Sec:Conclusions}.

\section{Bilinear field-oscillator coupling\label{Sec:MOF}}

\begin{figure}[!t]
\centering
\includegraphics[width=0.3\textwidth]{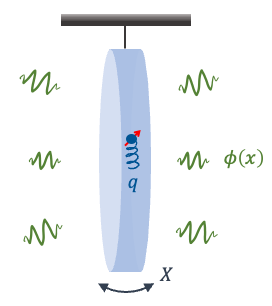}
\caption{Schematic representation of the system: The optical mirror of mass $M$ is confined by a harmonic potentials of frequency $\Omega$. Its displacement from the equilibrium postion is indicated by the coordinate $X$. The mirror interacts with a scalar field $\phi(x)$ via the mediation of the internal harmonic oscillator $q$, whose characteristic frequency is $\omega_0$, while its mass is $m$.
%As discussed in the text, the field can be expanded in the two independent sets of $(\pm)$ eigen-modes, whose spectrum is equal to $\omega_n = (2\pi n/L)$ $(n = 1,2,...)$, with $L$ the quantization length
}
\label{Fig:1}
\end{figure}

Let us consider the 1D system composed by a moving mirror of mass $M$, confined within a harmonic potential of frequency $\Omega$, interacting with the scalar field $\phi$. We indicate by $X$ the degrees-of-freedom corresponding to the mechanical displacement of the mirror. By following Ref.~\cite{MOF1}, we assume that this coupling is mediated by an internal degree-of-freedom (idf) of the mirror, whose dynamics is modelled as a harmonic oscillator $q$ of mass $m$ and frequency $\omega_0$. This system is sketched in Fig.~\ref{Fig:1}. By taking into account the microscopic dynamics of the mirror, this model is able to describe basic dielectric properties of optical mirrors, such as their transparency to radiation at high frequencies. This allows us overcome issues related to the standard theory of radiation pressure commonly used in optomechanics~\cite{Law-MirFieldInt-1995,Law-MirFieldInt-2011}, according to which the interaction results from (unphysical) boundary conditions that force the field to vanish at the location of the (ideal) mirror. Specifically, we model the interaction between the field and the idf in terms of a bilinear (i.e. linear in the corresponding variables) coupling. As discussed in Ref.~\cite{MOF1}, this theory generalizes and includes as subcases other popular models in the literature~\cite{Barton1,Barton2,Kardar1,Kardar2}. As such, the approach we pursue is more general and aims towards a more complete theory of opromechanics. For the sake of completeness, we briefly describe in this section the main features of the model we use, and show how the internal dynamics of the mirror determine its response to radiation. The content of this section closely follows Ref.~\cite{MOF1}, to which we direct the interested reader for more details.

The dynamics of the full system is described by the following action:
\begin{multline}
	S[X,q,\phi] = \int dt\,\bigg\{\,\frac{1}{2}\int dx\,\left[\bigg(\frac{\partial\phi}{c\partial t}\bigg)^2-\bigg(\frac{\partial\phi}{\partial x}\bigg)^2\right] + \frac{M}{2}\left(\dot{X}^2(t)-\Omega^2 X^2(t)\right) + \frac{m}{2}\Big(\dot{q}^2(t)-\omega_0 q^2(t)\Big) \\+ \lambda q(t) \phi(t,x)\delta(x-X)\bigg\}.
	\label{Eq:S_MOF}
\end{multline}
Throughout this paper we use the standard notation and indicate total time derivatives with dots over symbols, while $c$ is the speed of light. The first three terms in Eq.~\eqref{Eq:S_MOF} account for the free evolution of the field, of the mechanical oscillations of the mirror and of the idf, respectively, while the last term describes the bilinear coupling between the idf and the field. Notice that, in the latter term, the delta function enforces the interaction to take place at the position of the mirror, and thus couples implicitly all the degrees-of-freedom of the system. We indicated by $\lambda$ the interaction strength.
In the rest of this section, we discuss the response of the mirror to the field, in terms of its transmittivity and reflectivity properties. To this end, we freeze for the time being the mechanical motion and assume the is mirror at rest at the bottom of the confining potential. We use coordinates such that this equilibrium position is at $x=0$. The frequency-dependent reflection $R(\omega)$ and transmission $T(\omega)$ coefficients are obtained from the equations of motion for the idf and the field, which can be deduced by varying the action, with respect to $q$ and $\phi$. These take the form:
\begin{align}
	\bigg(\frac{\partial^2\phi}{c^2\partial t^2}-\frac{\partial^2\phi}{\partial x^2}\bigg) &= \lambda q(t) \delta(x),\label{Eq:phi1}\\
	m\ddot{q} + m \Omega^2 q &= \lambda \phi(0,t).\label{Eq:q1}
\end{align}
From Eq.~\eqref{Eq:phi1}, we infer that the first spatial derivative of the field is discontinuous at the mirror's position, but the field itself is continuous at the same point. To determine the spectral response of the mirror to radiation, let us consider the plane wave $\phi_\omega$ of frequency $\omega$, incident from the left and scattered at the mirror's position.
%with reflection $R(\omega)$ and transmission $T(\omega)$ coefficients.
The spatial structure of this field mode has the form:
\begin{equation}
	\phi_\omega(x,t) \sim e^{-i\omega t} [\theta(-x) \phi_\omega^L(x)+ \theta(x) \phi_\omega^R(x)],
\end{equation}
where $\theta(x)$ is the standard Heaviside function and we defined the field $\phi_\omega^L(x)$ on the left and on the field $\phi_\omega^R(x)$ on the right of the mirror, as:
\begin{align}
	\phi_\omega^L(x) &= e^{ikx} + R(\omega) e^{-ikx},\label{Eq:PhiL}\\
	\phi_\omega^R(x) &= T(\omega) e^{ikx}.\label{Eq:PhiR}
\end{align}
The field $\phi_\omega^L(x)$ is the linear combination of the right-moving incident plane wave and the left-moving reflected component, while $\phi_\omega^R(x)$ represents the transmitted component.
Since the mechanical fluctuations of the mirror are frozen, the interaction involves only the field and the idf, and is quadratic. Thus, in the steady state, the idf oscillates in time with the same frequency of the field: $q(t) = A(\omega) \exp(-i\omega t)$. The amplitude $A(\omega)$ of the idf oscillations is obtained by substituting this expression into Eq.~\eqref{Eq:q1}, together with the field evaluated at the location of the mirror: $\phi(t,0) = \phi_\omega^R(0)e^{-i\omega t} = T(\omega) e^{-i\omega t}$ (remember that the field is continuous across the mirror). The amplitude of idf oscillations are then obtained in the form:
\begin{equation}
	A(\omega) = \lambda\frac{T(\omega)}{m(\omega_0^2-\omega^2)}. \label{Eq:A}
\end{equation}
\begin{figure}[!t]
\centering
\includegraphics[width=0.6\textwidth]{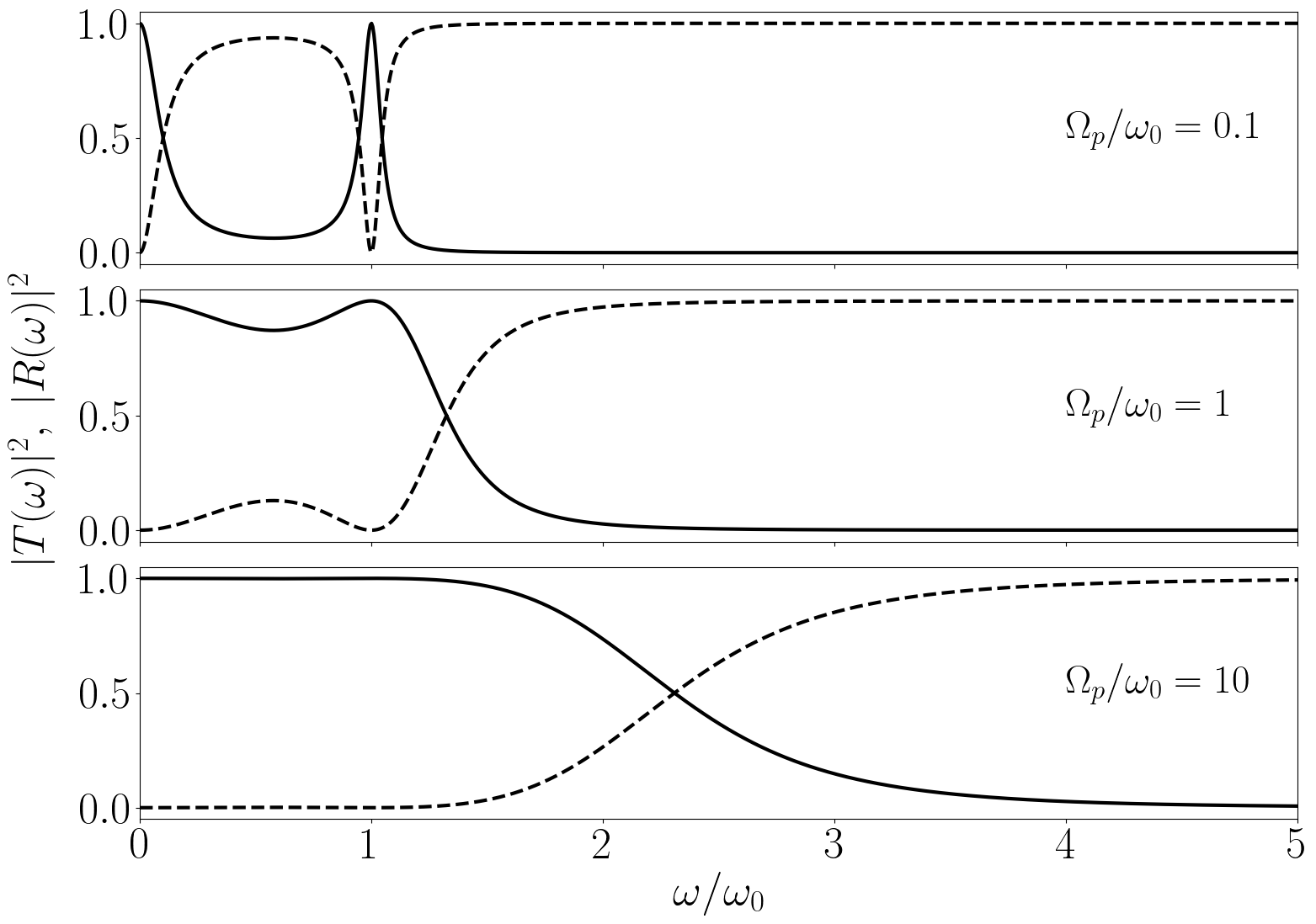}
\caption{Frequency dependence of the reflection (solid lines) and transmission (dashed lines) coefficients, as predicted by the field-idf bilinear interaction model. Each panel displays the results evaluated for different values of the ratio $\Omega_p/\omega_0$ between the plasma frequency and the oscillation frequency of the idf.}
\label{Fig:2}
\end{figure}
The continuity of the field and the discontinuity of its spatial derivative at the mirror's location, provide us with the two conditions needed to determine the frequency-dependent reflection $R(\omega)$ and transmission $T(\omega)$ coefficients. In particular, the jump of the spatial derivative of the field across the mirror is obtained by integrating the equation of motion Eq.~\eqref{Eq:phi1} over an infinitesimally small interval encompassing the mirror. These conditions give us the following set of equations:
\begin{align}
	\phi_\omega^L (0) &= 	\phi_\omega^R (0) ,\\
	\lim_{\epsilon\to 0}\int_{-\epsilon}^\epsilon dx\bigg(\frac{\partial^2\phi}{c^2\partial t^2}-\frac{\partial^2\phi}{\partial x^2}\bigg) &= -\frac{\partial \phi_\omega^R (t,0^+)}{\partial x} + \frac{\partial \phi_\omega^L(t,0^-)}{\partial x} = \lambda q(t).
\end{align}
Upon substitution of Eqs.~\eqref{Eq:PhiL} and \eqref{Eq:PhiR}, and given the steady state form for $q(t)$ given above, with the amplitude $A(\omega)$ as specified in Eq.~\eqref{Eq:A}, these conditions give us the following closed set of equations:
\begin{align}
	1+R(\omega) &= T(\omega),\\
	ik[1-R(\omega)] &= \bigg[ik+\frac{\lambda^2}{m(\omega_0^2-\omega^2)}\bigg] T(\omega),
\end{align}
from which the transmission and reflection coefficients are obtained in the form:
\begin{align}
	T(\omega) &= \frac{2m\omega(\omega_0^2-\omega^2)}{2m\omega(\omega_0^2-\omega^2)-i\lambda^2c},\\
	R(\omega) &= \frac{i\lambda^2 c}{2m\omega(\omega_0^2-\omega^2)-i\lambda^2 c}.
\end{align}
The squared modulo of these coefficients can be conveniently written by using the auxiliary function $F(\omega;\omega_0,\Omega_p) \equiv ({\omega}/{\Omega_p})[1-({\omega}/{\omega_0})^2]$, as:
\begin{align}
	|T(\omega)|^2 &= \frac{F^2(\omega;\omega_0,\Omega_p)}{1+F^2(\omega;\omega_0,\Omega_p)},\label{Eq:Tsq}\\
	|R(\omega)|^2 &= \frac{1}{1+F^2(\omega;\omega_0,\Omega_p)},\label{Eq:Rsq}
\end{align}
where $\Omega_p \equiv c\lambda^2/(2m\omega^2)$ is the \emph{plasma frequency} that characterizes the internal oscillations of the mirror. For illustrative purpose, Eqs.~\eqref{Eq:Tsq} and \eqref{Eq:Rsq} are plotted in Fig.~\ref{Fig:2}, for different values of the ratio $\Omega_p/\omega_0$. As evident from the figure, and discussed in details in Ref.~\cite{MOF1}, the relative value of the plasma frequency respect to the frequency of the idf specifies the operating regime of the mirror: In the case $\Omega_p/\omega_0\ll 1$ ($\Omega_p/\omega_0 = 0.1$ in Fig.~\ref{Fig:2}), the reflection coefficient is sharply peaked in a small interval around the idf frequency, so that the mirror reflects the incident radiation in a narrow bandwidth centred at $\omega_0$. In the opposite limit $\Omega_p/\omega_0\gg 1$  instead ($\Omega_p/\omega_0 = 10$ in the same figure), the mirror reflects the incident radiation over a much broader low-frequency bandwidth. These results demonstrate that the model is able to account for the partial reflectivity properties of real optical mirrors, and correctly captures its transparency to high-frequency radiation. We can take the characteristic frequency $\omega^*$ at which this transition happens as the one at which $|T(\omega^*)|^2 = |R(\omega^*)|^2 = 1/2$. In such conditions: $F^2(\omega^*) = 1$, inspection of which gives the scaling $\omega^*\sim \Omega_p^{1/3}$, in the limit $\Omega_p/\omega_0 \gg 1$.

\section{Hamiltonian for the moving mirror coupled to the field\label{Sec:MovingMirror}}

Having clarified the response of the mirror to radiation, that stem from the idf-field bilinear coupling, we move now to study the configuration in which the mirror oscillates within its trapping potential. In the rest of the paper we use the Hamiltonian formalism of quantum mechanics, and pursue an open quantum system approach to formulate an effective theory for the motional dynamics of the mirror, as it results from its interaction with the field via the mediation of the idf. Within this framework, field and idf represent the \emph{environment} for the mirror. The Hamiltonian that describes the dynamics of the whole system can be directly derived from the Lagrangian introduced in Eq.~\eqref{Eq:S_MOF}, and takes the form:
\begin{equation}
	\hat{H} = \hat{H}_{\rm M} + \hat{H}_{\rm I} + \hat{H}_{\rm F} + \hat{H}_{\rm int},\label{Eq:H}
\end{equation}
where we defined the Hamiltonians for the free evolution of the mirror (M), the idf (I) and the field (F), as:
\begin{align}
	\hat{H}_{\rm M} & = \frac{\hat{P}^2}{2M} + \frac{1}{2} M\Omega^2 \hat{X}^2,\label{Eq:H_M}\\
	\hat{H}_{\rm I} & = \frac{\hat{p}^2}{2m} + \frac{1}{2} m\omega_0^2 \hat{q}^2,\label{Eq:H_I}\\
	\hat{H}_{\rm F} & = \int{\frac{dx}{2}\,\Big[\hat{\pi}^2 + (\partial_x \hat{\phi})^2\Big]}.\label{Eq:H_F}
\end{align}
Here, $\hat{P}\equiv M d\hat{X}/dt$ and $\hat{p} \equiv m d\hat{q}/dt$ are the momenta conjugate to the mirror and idf oscillations, respectively, while $\hat{\pi}\equiv(1/c)(\partial_t \hat{\phi})$ is the momentum density conjugate to the scalar field. The Hamiltonian that describes the idf mediated interaction between the mirror and the field has the form:
\begin{equation}
	\hat{H}_{\rm int} = \lambda \int{ dx\, \hat{q} \hat{\phi}(x,t) \delta(x-\hat{X}) } = \lambda \hat{q} \hat{\phi}(x = \hat{X},t).\label{Eq:H_int}
\end{equation}
As noticed in the previous section, the delta function in Eq.~\eqref{Eq:H_int} enforces the interaction to take place at the position of the mirror, and thus couples all the degrees-of-freedom of the system.
For convenience, let us consider a finite quantization length $L$, with periodic boundary conditions, and expand the field in the $\{\cos(\kappa_n x),\, \sin(\kappa_n x)\}$ eigenbasis. Here, $\kappa_n \equiv 2\pi n/L$ (with $n \in \mathbb{N}$) are the wave vectors that are consistent with such boundary conditions. By using this basis, the field can be written in the form:
\begin{equation}
	\hat{\phi}(x,t) = \sum_{n=0}^{+\infty} \sqrt{\frac{2c^2}{L}}\Big[ \hat{q}_n^{-}(t) \cos(\kappa_n x) + \hat{q}_n^{+}(t) \sin(\kappa_n x)\Big],\label{Eq:F_Dec}
\end{equation}
with $\hat{q}_n^{\pm}(t)$ the time-dependent amplitudes of each mode. By using Eq.~\eqref{Eq:F_Dec}, it is straightforward to demonstrate that the free Hamiltonian of the field, Eq.~\eqref{Eq:H_F}, decomposes into the two independent $(\pm)$ sectors, as:
\begin{equation}
	\hat{H}_F = \hat{H}_F^{+} + \hat{H}_F^{-},
\end{equation}
with
\begin{equation}
	\hat{H}_F^{(\pm)} = \sum_{n=0}^{+\infty} \frac{1}{2} \bigg(\big[\hat{p}_n^{(\pm)}\big]^2 + \omega_n^2 \big[\hat{q}_{n}^{(\pm)}\big]^2 \bigg).\label{Eq:pm}
\end{equation}
We indicated by $\hat{p}_n^{(\pm)} \equiv {d\hat{q}}_n^{(\pm)}/dt$ the momenta conjugate to the field amplitudes $\hat{q}_n^{(\pm)}$.

In what follows, we make the usual approximation in optomechanics, that is we work in the small displacement limit and assume that the oscillations of the mirror around its equilibrium position are small compared to the wavelength of the optical modes that significantly interact with the mirror. This is a condition that is satisfied in typical optomechanical configurations \cite{Aspelmeyer_RMP}, as well as in experiments with trapped atoms or nanoparticles \cite{Isart-nanopart-2018}. We highlight that the model we use to describe the radiation pressure interaction is consistent with this assumption, since it correctly describes the partial reflectivity of the mirror, and thus its transparency for field modes of frequency above the plasma frequency. By working in this limit, we can expand the field in powers with respect to the small parameter $\epsilon\equiv \kappa_n{X}$, and write the interaction Hamiltonian in Eq.~\eqref{Eq:H_int}, up to second order, as:
\begin{equation}
	\hat{H}_{\rm int} = \lambda \hat{q}\Big[  \hat{\phi}(0,t) + \hat{X} \partial_x\hat{\phi}(0) + \frac{1}{2} \partial_x^2\hat{\phi}(0) \hat{X}^2\Big].\label{Eq:H_int_2}
\end{equation}
As discussed in the previous section, the first term in Eq.~\eqref{Eq:H_int_2} describes the direct interaction between the idf and the field, with the mirror at rest in its equilibrium position. The second and third terms instead describe the tripartite interaction involving the idf, the mirror and the field. By using the field decomposition in Eq.~\eqref{Eq:F_Dec}, Eq.~\eqref{Eq:H_int_2} can be written as:
\begin{equation}
	\hat{H}_{\rm int} = \hat{H}_{\rm int}^{(0)} + \hat{H}_{\rm int}^{(1)} + \hat{H}_{\rm int}^{(2)},
\end{equation}
with
\begin{align}
	\hat{H}_{\rm int}^{(0)} &= \lambda \hat{q} \hat{\phi}(0,t) = \bar{\lambda} \sum_{n=0}^{+\infty} \hat{q} \hat{q}_n^{-},\label{Eq:Hint_0}\\
	\hat{H}_{\rm int}^{(1)} &= \lambda \hat{q} \partial_x\hat{\phi}(0,t) \hat{X} = \bar{\lambda} \sum_{n=0}^{+\infty} \hat{q} \hat{q}_n^{+} (\kappa_n \hat{X}),\label{Eq:Hint_1}\\
	\hat{H}_{\rm int}^{(2)} &= \frac{\lambda}{2} \hat{q} \partial_x^2\hat{\phi}(0,t) \hat{X}^2 = -\frac{\bar{\lambda}}{2} \sum_{n=0}^{+\infty} \hat{q} \hat{q}_n^{-} (\kappa_n \hat{X})^2.\label{Eq:Hint_2}
\end{align}
The terms $\hat{H}_{\rm int}^{(i)}$, in Eqs.~\eqref{Eq:Hint_0}-\eqref{Eq:Hint_2}, are of $i$th-order in the perturbative parameter $\epsilon$, and we introduced the scaled interaction constant $\bar{\lambda}\equiv \lambda\sqrt{2c^2/L}$. To keep trace of the perturbative expansion, it is convenient to group terms in the full Hamiltonian as:
\begin{equation}
	\hat{H} = \hat{H}_0 + \epsilon \hat{H}_{\rm int}^{(1)} + \epsilon^2 \hat{H}_{\rm int}^{(2)},
	\label{Eq:H_eps}
\end{equation}
where, $\hat{H}_0 \equiv \hat{H}_{\rm M} + \hat{H}_{\rm I} + \hat{H}_{\rm F}^{+} + \hat{H}_{\rm F}^{-} + \hat{H}_{\rm int}^{(0)}$, and we explicitly indicated the power of the perturbative parameter $\epsilon$ in each term (this will be eventually set to the identity). Notice that the term $\hat{H}_{\rm int}^{(0)}$, that describes the idf-field bilinear coupling, is of zeroth order in the perturbative expansion with respect to small fluctuations of the mirror around its equilibrium position. This means that the finite reflectivity of the mirror to radiation is accounted for at any order of the perturbative expansion. The objective of the next section is to derive the master equation that describes the mechanical motion of the mirror, as it results due to its interaction with the field.

\section{Master equation for the moving mirror\label{Sec:ME}}

\subsection{Effective dynamics, noise and dissipation}
In the Schr\"odinger picture, the density operator $\hat{\chi}(t)$ of the system evolves in time according to the Liouville–von Neumann equation~\cite{Petruccione-book}:
\begin{equation}
	i\hbar\dot{\hat{\chi}}(t) = [\hat{H},\hat{\chi}(t)].
	\label{Eq:Chi_1}
\end{equation}
Since the interaction terms $\hat{H}_{\rm int}^{(1)}$ and $\hat{H}_{\rm int}^{(2)}$ are respectively cubic and quartic in the system variables, an exact analytical solution to Eq.~\eqref{Eq:Chi_1} cannot be found, in general. Aiming to an approximate solution, we pursue a perturbative method \cite{QBM2} and decompose the density operator as the sum of contributions of different order in $\epsilon$, that is: $\hat{\chi}(t) \equiv \hat{\chi}^{(0)}(t) + \epsilon \hat{\chi}^{(1)}(t) + \epsilon^2 \hat{\chi}^{(2)}(t) + \mathcal{O}(\epsilon^3)$. By using this expression, together with Eq.~\eqref{Eq:H_eps}, in Eq.~\eqref{Eq:Chi_1} and equating terms of the same order in $\epsilon$ on both sides, we obtain the following set of equations for $\hat{\chi}^{(i)}(t)$, up to second order:
\begin{align}
	i\hbar \dot{\hat{\chi}}^{(0)} (t)&= [\hat{H}_0 , \hat{\chi}^{(0)}(t)],\label{Eq:chi_0}\\
	i\hbar \dot{\hat{\chi}}^{(1)}(t) &= [\hat{H}_0 , \hat{\chi}^{(1)}(t)] + [\hat{H}_{\rm int}^{(1)} , \hat{\chi}^{(0)}(t)],\label{Eq:chi_1}\\
	i\hbar \dot{\hat{\chi}}^{(2)}(t) &= [\hat{H}_0 , \hat{\chi}^{(2)}(t)] + [\hat{H}_{\rm int}^{(1)} , \hat{\chi}^{(1)}(t)] + [\hat{H}_{\rm int}^{(2)} , \hat{\chi}^{(0)}(t)]\label{Eq:chi_2}.
\end{align}
These can be gathered together, giving:
\begin{equation}
	i\hbar\dot{\hat{\chi}} (t)= [\hat{H}_0 , \hat{\chi}(t)] + [(\hat{H}_{\rm int}^{(1)} + \hat{H}_{\rm int}^{(2)}) , \hat{\chi}^{(0)}(t)] + [\hat{H}_{\rm int}^{(1)} , \hat{\chi}^{(1)}(t)] + \mathcal{O}(\epsilon^3),
	\label{Eq:ME_1}
\end{equation}
As prescribed by Eq.~\eqref{Eq:chi_0}, the zeroth order term $\hat{\chi}^{(0)}(t)$ of the density operator evolves in time with the free Hamiltonian $\hat{H}_0$, while the first order term $\hat{\chi}^{(1)}(t)$ can be formally integrated from Eq.~\eqref{Eq:chi_1}. To this end, we work for convenience in the basis rotating with the free Hamiltonian $\hat{H}_{\rm 0}$, that is in the so-called interaction picture, in which the generic time-dependent operator $\hat{O}(t)$ is transformed as:
\begin{equation}
	\hat{O}_I(t) \equiv e^{i \hat{H}_0 (t-t_0)/\hbar} \, \hat{O}(t) \, e^{-i \hat{H}_0 (t-t_0)/\hbar},\label{Eq:IntPict}
\end{equation}
with $t_0$ the time at which we assume the interaction is switched on. Within this picture, states evolves in time only by effect of the interaction, and the first order term of the density matrix is formally integrated, as:
\begin{equation}
\hat{\chi}_I^{(1)} (t) = \frac{1}{i\hbar}\int_{t_0}^t ds \, [ \hat{H}_{{\rm int},I}^{(1)}(s), \hat{\chi}_I^{(0)} (s)]. \label{Eq:chi_I1}
\end{equation}
The master equation for the reduced density matrix of the mirror $\hat{\rho}_M$ is obtained by tracing Eq.~\eqref{Eq:ME_1} over the idf and the field degrees-of-freedom: $\hat{\rho}_M(t) \equiv {\rm Tr}_{+} {\rm Tr}_{I} {\rm Tr}_{-} \{\hat{\chi}(t)\}$. To this end, we make the usual assumption according to which all the degrees-of-freedom of the system are uncorrelated at the initial time $t=t_0$. This means that initially the density operator of the whole system factorizes as the Kronecker product of the density operators of each subsystem:
\begin{equation}
	\hat{\chi}(t=t_0) = \hat{\rho}_M^{(0)}(t_0) \otimes \hat{\rho}_I^{(0)}(t_0) \otimes \hat{\rho}_{+}^{(0)}(t_0) \otimes \hat{\rho}_{-}^{(0)}(t_0).
\label{Eq:Chi_T0}
\end{equation}
By inserting Eq.~\eqref{Eq:chi_I1} into Eq.~\eqref{Eq:ME_1}, and tracing over the field and the idf, we obtain the master equation for the mechanical motion of the mirror in the Redfield form \cite{Petruccione-book}:
\begin{multline}
	i\hbar\dot{\hat{\rho}}_M = [\hat{H}_{\rm M},\hat{\rho}_{\rm M}] + \frac{{\lambda}}{2} \langle \hat{q}(t) \partial_x^2\hat{\phi}(0,t) \rangle  [\hat{X}^2,\hat{\rho}_{\rm M}] \\+  \frac{\hbar\bar{\lambda}^2}{i\hbar}\int_{t_0}^t ds \,\Big\{ \nu_M(t,s) [\hat{X},[\hat{X}_I(s-t),\hat{\rho}_M(t)]] + i\mu_M(t,s) [\hat{X},\{ \hat{X}_I(s-t), \hat{\rho}_M(t)\}] \Big\}. \label{Eq:ME-2}
\end{multline}
The first term in Eq.~\eqref{Eq:ME-2} describes the unperturbed unitary evolution of the mirror within its confining potential, while the second term gives rise to a correction to the bare frequency of this motion, given by: $\Delta\Omega^2_1 = (\lambda/M) \langle \hat{q}\partial^2_x\hat{\phi}(0)\rangle$. Such a term is of second order in the pertubation theory, and results from tracing $\hat{H}_{\rm int}^{(2)}$ over the unperturbed total density matrix: ${\rm Tr}_{+} {\rm Tr}_{I} {\rm Tr}_{-}\{[\hat{H}_{\rm int}^{(2)},\hat{\chi}^{(0)}]\}$. The last term in Eq.~\eqref{Eq:ME-2} is also of second order, but is obtained instead from the trace: ${\rm Tr}_{+} {\rm Tr}_{I} {\rm Tr}_{-}\{[\hat{H}_{\rm int}^{(1)},\hat{\chi}^{(1)}]\}$. It accounts for memory effects that are encoded in the so-called fluctuation $\nu_M(t,s)$ and dissipation $\mu_M(t,s)$ kernels. These kernels account respectively for the noise and dissipation experienced by the mirror because of its coupling with the environment, and are defined in terms of symmetric and anti-symmetric combinations of the two-point correlation functions of the idf and the field. Specifically:
\begin{align}
	\nu_M (t,s) & = \frac{1}{2\hbar^2}\langle\{\hat{r}_I^{+}(t)\hat{q}_I(t),\hat{r}_I^{+}(s)\hat{q}_I(s)\}\rangle,\label{Eq:nuM}\\
	i\mu_M (t,s) & = \frac{1}{2\hbar^2}\langle[\hat{r}_I^{+}(t)\hat{q}_I(t),\hat{r}_I^{+}(s)\hat{q}_I(s)]\rangle,\label{Eq:muM}
\end{align}
where $\hat{r}^+ \equiv \sum_n \kappa_n \hat{q}_n^+$, and the two-time correlation functions are defined as:
\begin{align}
	&\big<\hat{q}_I(t) \hat{q}_I(s)\big> \equiv \text{Tr}_I\text{Tr}_{-}\{\hat{q}_I(t) \hat{q}_I(s)\, \hat{\rho}_{I}^{(0)}(t_0)\otimes\hat{\rho}_{-}^{(0)}(t_0)\},\label{Eq:Sigma}\\
	&\big<\hat{r}_I^{+}(t) \hat{r}_I^{+}(s)\big> \equiv \text{Tr}_{+}\{\hat{r}_I^{+}(t) \hat{r}_I^{+}(s) \,\hat{\rho}_{+}^{(0)}(t_0)\}.\label{Eq:Delta}
\end{align}
A first property that we can deduce from the definitions in Eqs.~\eqref{Eq:nuM} and \eqref{Eq:muM}, is that the noise kernel is an even function of time, $\nu_M(s,t) = \nu_M(t,s)$, while the dissipation kernel is an odd function, $\mu_M(s,t) = - \mu_M(t,s)$. Notice also that the correlations of the idf, Eq.~\eqref{Eq:Sigma}, depend on the state of the $(-)$ field's oscillators, due to the direct idf-field coupling described by the Hamiltonian $\hat{H}_{\rm int}^{(0)}$.
To make further progress and write the master equation in a form that helps a clearer physical interpretation of the dynamics of the mirror under the influence of its environment, we substitute in Eq.~\eqref{Eq:ME-2} the explicit expression for the time-dependent position operator $\hat{X}_I(t)$. This evolves in time according to the free Hamiltonian $\hat{H}_M$, and takes the form:
\begin{equation}
	\hat{X}_I(t) = \hat{X} \cos[\Omega (t-t_0)]+ \frac{\hat{P}}{M\Omega} \sin[\Omega (t-t_0)]. \label{Eq:Xt}
\end{equation}
By using Eq.~\eqref{Eq:Xt}, the master equation in Eq.~\eqref{Eq:ME-2} can finally be written in the form:
\begin{equation}
	i\hbar\dot{\hat{\rho}}_M(t) = [\hat{H}_{\rm M}^{\rm ren},\hat{\rho}_{\rm M}(t)] + i D_{PP}(t) [\hat{X},[\hat{X},\hat{\rho}_M(t)]] + i D_{XP}(t) [\hat{X},[\hat{P},\hat{\rho}_M(t)]] + \Gamma(t) [\hat{X},\{\hat{P},\hat{\rho}_M(t)\}].\label{Eq:ME-3}
\end{equation}
Here, we defined the renormalized mirror's Hamiltonian:
\begin{equation}
 	\hat{H}_{\rm M}^{\rm ren} = \frac{\hat{P}^2}{2M} + \frac{1}{2} M\Omega_{\rm ren}^2(t)\hat{X}^2,\label{Eq:HM_Ren}
\end{equation}
where: $\Omega_{\rm ren}^2(t) = \Omega^2 + \Delta\Omega^2_1(t) + \Delta\Omega^2_2(t)$ is the renormalized frequency of the mechanical vibrations. Beside the frequency shift $\Delta\Omega_1^2$ previously introduced, a further frequency shift $\Delta\Omega_2^2$ concurs in the definition of the renormalized frequency, that is defined as:
\begin{equation}
\Delta\Omega_2^2(t) = \frac{2\hbar\bar{\lambda}^2}{M} \int_{t_0}^t ds \cos\big[\Omega (s-t)\big] \mu_M(s,t).\label{Eq:Omega2}
\end{equation}
The other time-dependent coefficients appearing in the master equation take instead the form:
\begin{align}
	D_{PP}(t) &= -{\hbar\bar{\lambda}^2} \int_{t_0}^t ds \cos\big[\Omega (s-t)\big] \nu_M(s,t),\label{Eq:D_PP}\\
	D_{XP}(t) &= -\frac{\hbar\bar{\lambda}^2}{M\Omega} \int_{t_0}^t ds \,{\sin\big[\Omega (s-t)\big]} \nu_M(s,t),\label{Eq:D_XP}\\
	\Gamma(t) &= \frac{\hbar\bar{\lambda}^2}{M\Omega} \int_{t_0}^t ds \,{\sin\big[\Omega (s-t)\big]} \mu_M(s,t).\label{Eq:Gamma}
\end{align}
In Sec.~\ref{Sec:Reps}, we project the master equation in position basis and give also its phase space representation, showing the $D_{PP}(t)$ and $D_{XP}(t)$ are diffusion coefficients in the variables $p^2$ and $xp+px$, respectively, while $\Gamma(t)$ has the physical meaning of a damping coefficient. Importantly, these are defined in terms of the $\nu_M(t,s)$, $\mu_M(t,s)$ kernels, which carry information on the microscopic physical properties of the environment experienced by the mirror.

\subsection{Friction and diffusion coefficients in the stationary limit}
Let us assume that the interaction is switched on at the past infinity, that is we take the limit $t_0 \to -\infty$. At the finite time $t$, the system has thus reached the stationary regime and the noise and dissipation kernels become functions only of the time difference: $\lim_{t_0\to-\infty}\nu_M(s,t) \equiv \nu_M^{s}(s-t)$ (similarly for $\mu_M(s,t)$). In this stationary state, the noise $\nu_M^s(t)$ and dissipation $\mu_M^s(t)$ kernels are related by fluctuation-dissipation relations (FDRs), that we present in Sec.~\ref{Sec:NuMuM}. In turn, these relations allow us to find a direct connection between the asymptotic values of the diffusion $D_{PP}$ and friction $\Gamma$ coefficients. To find this connection, we perform the change of variables $\tau = t-s$ in Eqs.~\eqref{Eq:D_PP} and \eqref{Eq:Gamma}, and note that the integrand in these equations are even functions of time. This allows use to write the asymptotic values:
\begin{align}
	D_{PP}^s & \equiv D_{PP}(t_0\to -\infty) =  -\frac{\hbar\bar{\lambda}^2}{2} \int_{-\infty}^\infty d\tau \, e^{i\Omega\tau} \nu_M^s(\tau) \equiv  -\frac{\hbar\bar{\lambda}^2}{2} \tilde{\nu}_M^s(\Omega),\label{Eq:D_PP_Lim}\\
	\Gamma^s & \equiv \Gamma(t_0\to -\infty) = -i\frac{\hbar\bar{\lambda}^2}{2M\Omega} \int_{-\infty}^\infty d\tau \, e^{i\Omega\tau} \mu_M^s(\tau) \equiv  -i\frac{\hbar\bar{\lambda}^2}{2M\Omega} \tilde{\mu}_M^s(\Omega).\label{Eq:D_Gamma_Lim}
\end{align}
In the last equality in Eqs.~\eqref{Eq:D_PP_Lim} and \eqref{Eq:D_Gamma_Lim}, we introduced the Fourier transform of the noise and dissipation kernels, respectively. For the generic function of time $f(t)$, we define this as:
\begin{align}
	\tilde{f}(\omega) = \int_{-\infty}^\infty d\tau \, e^{i\omega\tau} f(\tau),
\end{align}
together with its inverse transform:
\begin{align}
	f(t) =\frac{1}{2\pi} \int_{-\infty}^\infty d\omega \, e^{-i\omega t} \tilde{f}(\omega).
\end{align}
Eqs.~\eqref{Eq:D_PP_Lim} and \eqref{Eq:D_Gamma_Lim} show that the asymptotic values of the $D_{PP}$ and $\Gamma$ coefficients are proportional to the spectral amplitude of the stationary noise and dissipation kernels $\tilde{\nu}_M^s(\Omega)$, $\tilde{\mu}_M^s(\Omega)$, evaluated at the frequency of the mechanical oscillations of the mirror. A similar simple relation does not hold for the coefficients $D_{XP}$ and $\Delta \Omega_2^2$, whose stationary values are instead sensitive to the whole frequency spectrum~\cite{MaiaNeto-PRA-2000,Dalvit-PRL-2000}.

The asymptotic values $D_{PP}^s$ and $\Gamma^s$ can be directly related to each other, by using the FDR derived in Sec.~\ref{Sec:NuMuM} (see Eq.~\eqref{Eq:Mirror_FDR}), that we report here for convenience: $\tilde{\nu}_M^s(\omega) = iz(\omega)\tilde{\mu}_M^s(\omega)$, with $z(\omega)\equiv\coth[\hbar\omega/(2k_BT)]$. By using this relation, along with Eqs.~\eqref{Eq:D_PP_Lim} and \eqref{Eq:D_Gamma_Lim}, we obtain:
\begin{equation}
	D_{PP}^s = M\Omega z(\Omega)\Gamma^s.\label{Eq:DsGs}
\end{equation}
Eq.~\eqref{Eq:DsGs} is a manifestation of the FRD, in a form that involves macroscopic signatures of the back-reaction, that are encoded in the diffusion and friction acting on the mirror.
In the low-temperature limit $\hbar\Omega\gg k_B T$ $(z(\Omega)\to 1)$, this relation reduces to $D_{PP}^s = M\Omega\Gamma^s$ while, in the opposite high-temperature limit $\hbar\Omega\ll k_B T$ $(z(\Omega)\to 2k_BT/(\hbar\Omega))$, it yields the standard Kubo form~\cite{QBM2,Caldeira-PhysA-1983} $D_{PP}^s = (2M k_B T/\hbar)\Gamma^s$.

\section{Noise and dissipation kernels for the mirror\label{Sec:NoiseDiss}}
In the previous section, we derived the master equation that governs the mechanical motion of the mirror interacting with the radiation field. We identified macroscopic effects of the back-reaction of the field in the friction and diffusion experienced by the mirror, whose strengths are characterized by the values of the coefficients $\Gamma(t)$, $D_{XX}(t)$ and $D_{XP}(t)$. Underlying these macroscopic effects are microscopic processes that account for the mirror-field interaction, that are encoded in the noise $\nu_M(t,s)$ and dissipation $\mu_M(t,s)$ kernels. Such microscopic quantities carry information on the physical properties of the environment, in terms of combinations of two-time correlators for the field and the idf, as defined in Eqs.~\eqref{Eq:Sigma} and \eqref{Eq:Delta}. In this section, we calculate the formal expressions for $\nu_M(t,s)$ and $\mu_M(t,s)$, by first giving the form for the two-time correlators of the $(+)$ modes of the field and the idf.

\subsection{Correlations for the (+) field oscillators\label{Sec:Corr+}}
The two-time correlation function for the $(+)$ field oscillators are easily calculated in the assumption of initial thermal state of the system. Omitting the $(+)$ superscripts for ease in what follows, we have:
\begin{equation}
	\big<\hat{r}_I(t) \hat{r}_I(s)\big> = \sum _{k,j=0}^{+\infty} \kappa_k \kappa_j \big<\hat{q}_{k,I}(t) \hat{q}_{j,I}(s)\big>. \label{Eq:R}
\end{equation}
By inserting into \eqref{Eq:R} the expression for the time-dependent $(+)$ field amplitudes $\hat{q}_{k,I}(t)$, which evolve according to the free Hamiltonian $\hat{H}_F^+$, as:
\begin{equation}
	\hat{q}_{k,I}(t) = \hat{q}_k \cos[\omega_k (t-{t_0})] +\frac{\hat{p}_k}{\omega_k}\sin[\omega_k (t-{t_0})],
\end{equation}
we obtain:
\begin{multline}
	\big<\hat{r}_I(t) \hat{r}_I(s)\big> = \sum_{k=0}^{+\infty}\kappa_k^2 \Big\{\big<\hat{q}_k^2\big> \cos[\omega_k (t-{t_0})]\cos[\omega_k (s-{t_0})] + \frac{\big<\hat{q}_k \hat{p}_k\big>}{\omega_k} \cos[\omega_k (t-{t_0})]\sin[\omega_k (s-{t_0})] \\
	+\frac{\big<\hat{p}_k \hat{q}_k\big>}{\omega_k} \sin[\omega_k (t-{t_0})]\cos[\omega_k (s-{t_0})] + \frac{\big<\hat{p}_k^2\big>}{\omega_k^2} \sin[\omega_k (t-{t_0})]\sin[\omega_k (s-{t_0})]\Big\}.
\end{multline}
For a thermal state: $\big<\hat{q}_k \hat{p}_k\big> = \big<\hat{p}_k \hat{q}_k\big>^*  = i\hbar/2$, $\big<\hat{q}_k^2\big> = (\hbar/2\omega_k)(2n_k+1)$, $\big<\hat{p}_k^2\big> = (\hbar\omega_k/2)(2n_k+1)$, with $n_k$ the number of excitations in the $k$-th mode, and $(2n_k+1) = \coth(\beta\hbar\omega_k/2) \equiv z(\omega_k) $. This gives the sought expression for the correlator:
\begin{equation}
	\big<\hat{r}_I(t) \hat{r}_I(s)\big> = \hbar \big[ \nu_{+}(t-s) + i \mu_{+}(t-s)\big],
	\label{Eq:rr}
\end{equation}
where
\begin{align}
	\nu_{+}(t) = \sum_{k=0}^{+\infty} \frac{\omega_k}{2c^2}  z_k \cos(\omega_k t),\\
	\mu_{+}(t) = -\sum_{k=0}^{+\infty} \frac{\omega_k}{2c^2} \sin(\omega_k t),
\end{align}
are the fluctuation and dissipation kernels pertaining to the $(+)$ bath of field oscillators. As usual, these are related to the symmetric and anti-symmetric two-time correlators, as:
\begin{align}
	\langle \{ \hat{r}_I(t) , \hat{r}_I(0)\}\rangle &= 2\hbar \,\nu_{+}(t),\\
	\langle [ \hat{r}_I(t) , \hat{r}_I(0)]\rangle &= 2i\hbar \,\mu_{+}(t).
\end{align}
As expected, the two-time correlator $\langle\hat{r}_I(t)\hat{r}_I(s)\rangle$ depends only on the time difference $t-s$, since the field is in thermal equilibrium and thus stationary. It is useful to take the continuum limit $L\rightarrow +\infty$, and turn sums over modes into integrals over frequencies, according to the prescription: $\sum_k = \int d\omega \mathcal{D}(\omega)$, where $\mathcal{D}(\omega) = \mathcal{D} = c L/2\pi$ is the frequency density of states, that is constant in the one-dimensional case here considered. The qualitative behaviour of higher dimensional configurations can be captured by inserting into these integrals the corresponding frequency dependence of the density of states. In such a limit, the noise and dissipation kernels for the $(+)$ field oscillators can be written in the form:
\begin{align}
	\nu_{+}(t) = \mathcal{D}\int_{-\infty}^{+\infty}d\omega\,\frac{\omega}{4c^2}z(\omega) e^{-i\omega t} \equiv \frac{L}{2\pi}\int_{-\infty}^{+\infty}d\omega\,\, \tilde{\nu}_{+}(\omega) e^{-i\omega t},\\
		\mu_{+}(t) = \mathcal{D}\int_{-\infty}^{+\infty}d\omega\,\frac{(-i\omega)}{4c^2} e^{-i\omega t} \equiv \frac{L}{2\pi}\int_{-\infty}^{+\infty}d\omega\,\, \tilde{\mu}_{+}(\omega) e^{-i\omega t},
\end{align}
where
\begin{align}
	\tilde{\nu}_{+}(\omega) & = z(\omega)\frac{\omega}{4c},\label{Eq:nu+}\\
	\tilde{\mu}_{+}(\omega) & = -i\frac{\omega}{4c},\label{Eq:mu+}
\end{align}
can be identified (up to a factor $L$) as the Fourier transforms of the noise and dissipation kernels generated by the $(+)$ bath of field oscillators. By comparing Eqs.~\eqref{Eq:nu+} and \eqref{Eq:mu+}, we infer that these satisfy the standard FDR: $\tilde{\nu}_{+}(\omega) = i z(\omega) \tilde{\mu}_{+}(\omega) $.

\subsection{Correlations for the  microscopic internal oscillator\label{Sec:Corr-q}}
Calculating the two-time correlations for the idf is more involved because of its direct coupling with the $(-)$ field oscillators. Such an interaction is described by the Hamiltonian: $\hat{H}_{I,(-)} \equiv \hat{H}_I + \hat{H}_{-} + \hat{H}_{\rm int}^{-}$, which gives rise to the following set of equations for the time evolution of the idf and $(-)$ field modes amplitudes:
\begin{align}
	\ddot{\hat{q}}_I + \omega_0^2 \hat{q}_I &= - \frac{\bar{\lambda}}{m}\sum_{n=0}^{+\infty} \hat{q}_{n,I}^{-},\label{Eq:q}\\
	\ddot{\hat{q}}_{n,I}^{-} + \omega_n^2 \hat{q}_{n,I}^{-} &= - \bar{\lambda}\hat{q}_I. \label{Eq:qn}
\end{align}
Since we are interested in the correlation function of the internal oscillator, we first solve Eq.~\eqref{Eq:qn} for the field amplitudes, in terms of a combination of the corresponding homogeneous $(\hat{q}_{n,I}^-)_h$ and particular $(\hat{q}_{n,I}^-)_p$ solutions:
% $\hat{q}_{n,I}^-(t) = (\hat{q}_{n,I}^-)_h (t)+ (\hat{q}_{n,I}^-)_p (t)$
\begin{equation}
	\hat{q}_{n,I}^-(t) = (\hat{q}_{n}^-)_h (t)+ (\hat{q}_{n}^-)_p (t), \label{Eq:qn_full}
\end{equation}
with:
\begin{align}
	(\hat{q}_{n,I}^-)_h (t) &= \hat{q}_n^{-}\cos[\omega_n (t-{t_0})] + \hat{p}_n^- \frac{\sin[\omega_n (t-{t_0})]}{\omega_n},\label{Eq:qn_h}\\
	(\hat{q}_{n,I}^-)_p (t) &=	\int_{t_0}^{+\infty} ds\, G_n^-(t-s)(-\bar{\lambda}\hat{q}_I(s)).\label{Eq:qn_p}
\end{align}
In Eq.~\eqref{Eq:qn_h}, $\hat{q}_n^{-}$, $\hat{p}_n^{-}$ are the initial amplitude and conjugate momentum of the field's modes, respectively. In Eq.~\eqref{Eq:qn_p} instead, $G_n^-(t) = \theta(t)g_n^-(t)$ with $g_n^-(t) = {\sin(\omega_n t)}/{\omega_n}$, is the retarded Green function for Eq.~\eqref{Eq:qn}, with vanishing initial conditions. By using Eqs.~\eqref{Eq:qn_full}-\eqref{Eq:qn_p}, the equation governing the time evolution of the idf, that includes the back-reaction from the field, takes the form:
\begin{equation}
	\ddot{\hat{q}}_I + \omega_0^2 \hat{q}_I + \frac{2}{m} \int_{t_0}^{t} \,ds\,\mu_{-}(t-s)\hat{q}_I(s) = \frac{\hat{f}_{-}(t)}{m}.\label{Eq:q-1}
\end{equation}
This equation describes the dynamics of the idf subjected to the influence of the $(-)$ bath of field oscillators. This influence appears in the form of an external noise driving the oscillator, here indicated by $\hat{f}_{-}(t)$, and a non-local damping, generated by the memory kernel $\mu_-(t)$. These reads as:
\begin{align}
	\mu_{-}(t) &= - \frac{\bar{\lambda}^2}{2}\sum_{n=0}^{+\infty} g_n^-(t),\\
	\hat{f}_-(t) &= -\bar{\lambda}\sum_{n=0}^{+\infty}\Big\{\hat{q}_n^{-}\cos[\omega_n (t-t_0)] + \frac{\hat{p}_n^{-}}{\omega_n} \sin[\omega_n (t-t_0)]\Big\}.
\end{align}
As usual, noise and dissipation induced by the $(-)$ bath of field oscillators must be related by FDRs. Specifically, these relate the dissipation kernel $\mu_-(t)$ to the corresponding noise $\nu_-(t)$ kernel. The two are defined in terms of the symmetric and anti-symmetric correlation function of the force $\hat{f}_-(t)$, as:
\begin{align}
	\langle \{\hat{f}_-(t), \hat{f}_-(s) \}\rangle & = 2\hbar \,\nu_{-}(t-s) = 2\hbar\, \sum_{n=0}^{+\infty}\frac{\bar{\lambda}^2}{2\omega_n}z(\omega_n)\cos[\omega_n(t-s)],\\
	\langle [\hat{f}_-(t), \hat{f}_-(s)] \rangle & = 2i\hbar \,\mu_{-} (t-s).
\end{align}
By taking the continuum limit:
\begin{align}
	\nu_{-}(t) & = \mathcal{D}\int_{-\infty}^{+\infty} d\omega\, \frac{\bar{\lambda}^2}{4\omega}z(\omega) e^{-i\omega t} \equiv \frac{1}{2\pi}\int_{-\infty}^{+\infty} d\omega \,\tilde{\nu}_-(\omega) e^{-i\omega t},\\
	\mu_{-}(t) & = \mathcal{D} \int_{-\infty}^{+\infty}d\omega \left(\frac{-i\bar{\lambda}^2}{4\omega}\right) e^{-i\omega_k t} \equiv \frac{1}{2\pi} \int_{-\infty}^{+\infty} d\omega \,\tilde{\mu}_-(\omega) e^{-i\omega t},
\end{align}
we infer that the Fourier transforms of the noise and dissipation kernels for the $(-)$ bath of field oscillators, have the form:
\begin{align}
	\tilde{\nu}_{-}(\omega) &= z(\omega)\frac{\lambda^2c^3}{2\omega}, \\
	\tilde{\mu}_{-}(\omega) &= -i\frac{\lambda^2c^3}{2\omega}.
\end{align}
As expected, these satisfy the standard FDR relation: $	\nu_{-}(\omega) = i z(\omega) \mu_{-}(\omega)$.

Eq.~\eqref{Eq:q-1} can as well be formally solved in terms of a linear combination of the corresponding homogeneous solution $(\hat{q}_I)_h(t)$, which carries information of the initial state of the idf, and particular solution that propagates the effect of the external noise. The latter is written as the convolution of the retarded Green's function for Eq.~\eqref{Eq:q-1}, that we indicate as $G_q(t) = \theta(t) g_q(t)$, with the forcing term $\hat{f}_-(t)$. The solution can thus be formally written:
\begin{equation}
	\hat{q}_I(t) = (\hat{q}_I)_h(t) + \frac{1}{m}\int_{t_0}^{+\infty} ds\,G_q(t-s)\hat{f}_-(s).
\end{equation}
Calculating the explicit expression of the Green function $G_q(t)$ is not trivial, because of the non-Markovian nature of the system at hand, and goes beyond the scopes of this work.
In the assumption of idf and field initially uncorrelated, the two-time correlations of the idf is finally written in the form:
\begin{equation}
	\langle \hat{q}_I(t)\hat{q}_I(s)\rangle = \langle \hat{q}_h(t)\hat{q}_h(s)\rangle + \frac{1}{m^2}\int_{t_0}^{+\infty} d\tau \int_{t_0}^{+\infty} d\xi\, G_q(t-\tau)\langle \hat{f}_-(\tau)\hat{f}_-(\xi) \rangle  G_q(s-\xi),\label{Eq:qq}
\end{equation}
where
\begin{equation}
	\langle \hat{f}_-(\tau)\hat{f}_-(\xi) \rangle = \hbar[\nu_-(\tau-\xi) + i \mu_-(\tau-\xi)].
\end{equation}

\subsection{Structure of the noise and dissipation kernels and relation with the dynamical Casimir effect\label{Sec:NuMuM}}

In the previous sections, we obtained the two-point correlation functions for the $(+)$ field oscillators and the idf, given in Eqs.~\eqref{Eq:rr} and \eqref{Eq:qq}, respectively. By using these results, together with the definitions in Eqs.~\eqref{Eq:nuM} and \eqref{Eq:muM}, we are now in the position to give the formal expressions for the fluctuation and dissipation kernels of the mirror.
By noting that
$\langle \{\hat{q}_h(t),\hat{q}_h(s)\}\rangle = \langle \{\hat{q}_h(t),\hat{q}_h(s)\}\rangle^*$ and $\langle [\hat{q}_h(t),\hat{q}_h(s)]\rangle = -\langle [\hat{q}_h(t),\hat{q}_h(s)]\rangle^*$, these can be written as:
\begin{multline}
	\nu_M(t,s)= \frac{1}{\hbar}\Big[\nu_{+}(t-s) \text{Re}\langle (\hat{q}_I)_h(t),(\hat{q}_I)_h(s)\rangle - \mu_{+}(t-s)\text{Im}\langle (\hat{q}_I)_h(t),(\hat{q}_I)_h(s)\rangle\Big] \\
				+\frac{\bar{\lambda}^2}{m^2}\Big\{\nu_{+}(t-s)[G_q * \nu_{-} * G_q](t-s) - \mu_{+}(t-s)[G_q * \mu_{-} * G_q](t-s)\Big\},
\label{Eq:nuFull}
\end{multline}
and
\begin{multline}
	\mu_M(t,s) = \frac{1}{\hbar}\Big[\nu_{+}(t-s) \text{Im}\langle (\hat{q}_I)_h(t),(\hat{q}_I)_h(s)\rangle + \mu_{+}(t-s)\text{Re}\langle (\hat{q}_I)_h(t),(\hat{q}_I)_h(s)\rangle\Big] \\
				 +\frac{\bar{\lambda}^2}{m^2}\Big\{\mu_{+}(t-s)[G_q * \nu_{-} * G_q](t-s) + \nu_{+}(t-s)[G_q * \mu_{-} * G_q](t-s)\Big\}.\label{Eq:muFull}
\end{multline}
Here, we defined the two-time convolutions $[G_q * \nu/\mu_{-} * G_q](t-s)\equiv\int_{t_0}^{+\infty} d\tau \int_{t_0}^{+\infty} d\xi\, G_q(t-\tau) \nu/\mu_{-}(\tau-\xi) G_q(s-\xi)$. Eqs.~\eqref{Eq:nuFull} and \eqref{Eq:muFull} are the second main result of this paper, along with their frequency representation given in what follows. They carry a clear physical meaning:
Fluctuations in the mirror result from combinations of fluctuations in the $(+)$ modes of the field and fluctuations in the idf. The latter could be either encoded in the initial state of internal oscillator itself (first bracket in Eqs.~\eqref{Eq:nuFull} and \eqref{Eq:muFull}) or could originate from the $(-)$ bath of field oscillators and then propagated by the idf (second bracket in the same equation).
By taking the limit $t_0\to-\infty$, the contribution due to the homogeneous solution $\langle(\hat{q}_I)_h(t)(\hat{q}_I)_h(s)\rangle$ vanishes because of the damping introduced by the interaction with the $(-)$ bath of field oscillators. In this limit, the evolution becomes stationary, and only the second bracket in Eqs.~\eqref{Eq:nuFull} and  \eqref{Eq:muFull} is left. Fluctuation-dissipation relations for the noise and damping experienced by the mechanical oscillator are derived upon Fourier transforming $\nu_M(t)$ and $\mu_M(t)$ in the stationary regime. In the frequency domain these take the form:
\begin{align}
	\tilde{\nu}_M^s(\omega) &= \frac{2c^2{\lambda}^2}{m^2}\int_{-\infty}^{\infty}\frac{d\omega'}{2\pi}\Big\{\tilde{\nu}_{+}(\omega-\omega')\tilde{G}_q(\omega') \tilde{\nu}_{-}(\omega') \tilde{G}_q(-\omega') - \tilde{\mu}_{+}(\omega-\omega')\tilde{G}_q(\omega') \tilde{\mu}_{-}(\omega') \tilde{G}_q(-\omega')\Big\},\\
	\tilde{\mu}_M^s(\omega) &= \frac{2c^2{\lambda}^2}{m^2}\int_{-\infty}^{\infty}\frac{d\omega'}{2\pi}\Big\{\tilde{\mu}_{+} (\omega-\omega') \tilde{G}_q(\omega') \tilde{\nu}_{-}(\omega') \tilde{G}_q(-\omega') + \tilde{\nu}_{+}(\omega-\omega')\tilde{G}_q(\omega') \tilde{\mu}_{-}(\omega') \tilde{G}_q(-\omega')\Big\}.
\end{align}
Written in the frequency domain, the fluctuation and dissipation kernels reveal an intriguing physical interpretation. Fluctuations experienced by the mirror at frequency $\omega$ are induced by fluctuations at frequency $\omega_+ = \omega-\omega'$ in the $(+)$ sector of the field, and fluctuations at frequency $\omega_- = \omega'$ in the $(-)$ sector, such that the resonance condition $\omega = \omega_+ + \omega_-$ is verified. This suggests that fluctuations acting on the mirror at a certain frequency are generated by the emission of couple of photons in different sectors of the field, whose frequencies sum up to resonance. This is an indication that the dynamical Casimir effect is at the origin of noise experienced by the mirror. Interestingly, the result in Eq.~\eqref{Eq:D_Gamma_Lim} shows that the asymptotic values of the friction coefficient $\Gamma^s$ is sensitive to fluctuations at the frequency of the mechanical vibrations $\Omega$, making thus explicit the connection between damping induced onto the mirror and the creation of photon pairs.

By using the FDR for the $(\pm)$ field oscillators obtained in the previous sections, $\tilde{\nu}_{\pm}(\omega) = i z(\omega) \tilde{\mu}_{\pm}(\omega)$, eqs.~\eqref{Eq:nuFull} and \eqref{Eq:muFull} can be opportunely rearranged in the form:
\begin{align}
	\tilde{\nu}_M^s(\omega) &= \frac{2c^2{\lambda}^2}{m^2}\int_{-\infty}^{\infty}\frac{d\omega'}{2\pi}\Big\{[z(\omega-\omega')z(\omega')+1]\tilde{\mu}_{+}(\omega-\omega')\tilde{\mu}_{-}(\omega')G_q(\omega')G_q(-\omega')\Big\}\nonumber\\
	& \equiv \int_{-\infty}^{\infty}\frac{d\omega'}{2\pi} n(\omega,\omega'),\\
	\tilde{\mu}_M^s(\omega) &= i\frac{2c^2{\lambda}^2}{m^2}\int_{-\infty}^{\infty}\frac{d\omega'}{2\pi}\Big\{[z(\omega')+z(\omega-\omega')]\tilde{\mu}_{+} (\omega-\omega') \tilde{\mu}_{-}(\omega') G_q(\omega')  G_q(-\omega')\Big\}\nonumber\\
	& \equiv i\int_{-\infty}^{\infty}\frac{d\omega'}{2\pi} m(\omega,\omega').
\end{align}
From these expressions, we deduce that the integrand functions $n(\omega,\omega')$, $m(\omega,\omega')$ are related as $	n(\omega,\omega') = - z(\omega) m(\omega,\omega')$, from which
we finally obtain the sought FDR for the mirror, in the standard form:
\begin{equation}
	\tilde{\nu}_M^s (\omega) = \int_{-\infty}^{\infty}\frac{d\omega'}{2\pi} n(\omega,\omega')  = - \int_{-\infty}^{\infty}\frac{d\omega'}{2\pi} z(\omega) m(\omega,\omega') = i z(\omega) \tilde{\mu}_M^s (\omega). \label{Eq:Mirror_FDR}
\end{equation}

\section{Position and Wigner phase space representations\label{Sec:Reps}}

A clearer physical understanding of the dynamical effects induced onto the mirror by the DCE emission is achieved by projecting the master equation for the mirror, Eq.~\eqref{Eq:ME-3}, in the position basis and in phase space. We start by addressing the former task. In position basis, the reduced density matrix is written as $\hat{\rho}_M(x,x') \equiv \Melement{x}{x'}{\hat{\rho}_M}$ and the master equation takes the form \cite{QBM1,QBM2,QBM3}
\begin{multline}
	i\hbar\frac{\partial\rho_M(x,x';t)}{\partial t} = \bigg[-\frac{\hbar^2}{2M}\bigg(\frac{\partial^2}{\partial x^2}-\frac{\partial^2}{\partial x'^2}\bigg) + \frac{1}{2} M \Omega_{\rm ren}^2(t)(x^2-x'^2) + i D_{PP}(t)(x-x')^2\\
	+ \hbar D_{XP}(t)(x-x')\bigg(\frac{\partial}{\partial x}+\frac{\partial}{\partial x'}\bigg)-i\hbar\Gamma(t) (x-x')\bigg(\frac{\partial^2}{\partial x^2}-\frac{\partial^2}{\partial x'^2}\bigg)\bigg]\rho_M(x,x';t).
\label{Eq:ME_x}
\end{multline}
The first two terms on the right-hand-side of Eq.~\eqref{Eq:ME_x} account for the unitary evolution generated by the renormalized Hamiltonian, that is by the term $[\hat{H}_{\rm M}^{\rm ren},{\rho}_{\rm M}]$. The third and fourth terms instead, are responsible for quantum decoherence. Specifically, the third term suppress spatial coherence that account for long-range correlations between spatially separated components of the wave-function of the mirror. The fourth term suppresses instead mixed spatial-momentum coherence, that is correlations between components in the wave-function of the mirror, separated both in position and momentum. Finally, the last term describes dissipation effects induced by the environment onto the mirror, with $\Gamma(t)$ being the corresponding friction coefficient.
For completeness, we also give the phase space representation of the master equation. To this end, we introduce the Wigner quasi-probability distribution $W_M(x,p)$, that is a semi-classical representation of the reduced density matrix, in the phase space. This is defined as partial Fourier transform of the density matrix in the position basis, that is:
\begin{equation}
	W_M(x,p) = \frac{1}{2\pi\hbar} \int_{-\infty}^{+\infty} dy\, e^{ipy/\hbar}\,\rho_M\bigg(x-\frac{y}{2},x+\frac{y}{2}\bigg).
\label{Eq:WignerDef}
\end{equation}
By using this definition, the Eq.~\eqref{Eq:ME_x} is transformed into the corresponding equation for the Wigner distribution, taking the form:
\begin{equation}
	\frac{\partial W(x,p;t)}{\partial t} = \bigg[-\frac{p}{M}\frac{\partial}{\partial x} + M\Omega_{\rm ren}^2(t)\, x\frac{\partial}{\partial p} - \hbar D_{PP}(t) \frac{\partial^2}{\partial p^2} + \hbar D_{XP}(t) \frac{\partial^2}{\partial x\partial p} + 2\Gamma(t) \frac{\partial}{\partial p} p \bigg] W(x,p;t).
\label{Eq:ME_Wigner}
\end{equation}
The first two terms in the right-hand-side of Eq.~\eqref{Eq:ME_Wigner} give the classical Poisson bracket $\{H_{\rm M}^{\rm ren},W_M\} \equiv (\partial H_{\rm M}^{\rm ren}/\partial x) (\partial W_M/\partial p) - (\partial H_{\rm M}^{\rm ren}/\partial p) (\partial W_M/\partial x)$, that is the phase space representation of the Hamiltonian evolution. The third and fourth terms describe instead diffusion, respectively in the variables $p^2$ and $xp$, with $D_{PP}(t)$ and $D_{XP}(t)$ the corresponding diffusion coefficients. Diffusion is thus the physical process that accounts for decoherence in phase space. Finally, the last term is a drift term in momentum, that is damping. 

Decoherence effects generated by the interaction with the field are responsible for driving the quantum state of the mirror towards classicality~\cite{zurek1991decoherence}. To explicitly see this effect, let us consider for example the mirror prepared in the coherent superposition state $\psi_M(x)$ of minimum-uncertainty wave-packets, localized at opposite positions $x=\pm x_0/2$, and carrying opposite momentum $p=\pm p_0/2$. Such a state is described by the wave-function
\begin{equation}
	\psi_M(x) = \frac{1}{\sqrt{2}}[\varphi(x;{x_0}/{2},{p_0}/{2}) + \varphi(x;-{x_0}/{2},-{p_0}/{2})],\label{Eq:Superp}
\end{equation}
with
\begin{equation}
	\varphi(x;{x_0},{p_0}) = \frac{1}{(\pi\delta^2)^{1/4}} \exp\bigg[{-\frac{(x-x_0)^2}{2\delta^2}}+i\frac{p_0}{\hbar}x\bigg].
\end{equation}
An approximate value of the time scales $t_{XX}$ and $t_{XP}$ by which the system looses respectively spatial and spatial-momentum coherence, can be obtained by considering the stationary values of the diffusion coefficients in Eq.~\eqref{Eq:ME_x}. This allows us to estimate: $t_{XX} \approx {\hbar}/{(D_{PP}^s x_0^2)}$ and $t_{XP} \approx {\hbar}/{(D_{XP}^s x_0 p_0)}$, meaning that the more non-classical is the state of the mirror, that is the larger is the $x_0$ and $p_0$ separation in phase space of the wave-packets, the faster the mirror undergoes decoherence. The same results can of course be obtained by working in the Wigner representation. In this case, the coherent superposition in Eq.~\eqref{Eq:Superp} is described by the Wigner function
\begin{equation}
	W_M(x,p) = \frac{1}{2}\Big\{W_M(x,p)[x_0/2,p_0/2]+W_M(x,p)[-x_0/2,-p_0/2]+W_M(x,p)[0,0]\cos[(p x_0 - p_0 x)/\hbar]\Big\},\label{Eq:W_Superp}
\end{equation}
where
\begin{equation}
	W(x,p)[x_0,p_0] = \frac{1}{\pi\hbar} \exp\bigg[{-\frac{(x-x_0)^2}{2\delta^2}}\bigg] \exp\bigg[{-\frac{(p-p_0)^2}{\hbar^2}\delta^2}\bigg],
\end{equation}
is the Wigner distribution for the minimum uncertainty wave-packet $\varphi(x;x_0,p_0)$ localized at position $x_0$, and carrying momentum $p_0$. The quantum nature of the superposition state is witnessed by the last, oscillatory term in Eq.~\eqref{Eq:W_Superp}, that makes the Wigner function attaining negative values. Diffusion terms in Eq.~\eqref{Eq:ME_Wigner} act in such a way to wash out this oscillating component, at a rate given by the decoherence times given above.

A relation between the relaxation $t_R\approx 1/\Gamma^s$ and decoherence $t_{XX}$ time scales can be inferred by using Eqs.~\eqref{Eq:D_PP_Lim},\eqref{Eq:D_Gamma_Lim} and the FDR for $\nu_M(\omega)$ and $\mu_M(\omega)$ in Eq.~\eqref{Eq:Mirror_FDR}. Specifically:
\begin{equation}
	\frac{t_{XX}}{t_R} = \frac{\hbar\Gamma^s}{D_{PP}^s x_0^2} = \frac{i\hbar}{M\Omega D_{PP} x_0^2} \frac{\mu_M^s(\Omega)}{\nu_M^s(\Omega)} = \frac{\hbar}{M\Omega D_{PP} x_0^2z(\Omega)}.\label{Eq:tXX_tR}
\end{equation}
As expected, this result shows that for macroscopic massive objects the decoherence happens at a much shorter time scale than relaxation, making extremely hard to see quantum effects in the macroscopic world. Moreover, in the low-temperature limit: $\beta\hbar\Omega \gg 1$, the times scale ratio reduces to $t_{XX}/t_R \to \hbar/(M\Omega D_{PP} x_0^2)$, while in the opposite limit $\beta\hbar\Omega \ll 1$, $t_{XX}/t_R \to \hbar^2/(2k_B T M D_{PP} x_0^2)$. This makes explicit the role of the mechanical and temperature energy scales in the decoherence process.
Notice that similar relations between $t_{XP}$ and $t_R$ cannot be obtained because of the absence of a direct relation between $\Gamma^s$ and $D_{XP}^s$.

\section{Conclusions\label{Sec:Conclusions}}
In this work, we studied the effects of the back-reaction from a quantum field onto the dynamics of a moving mirror. We made use of a microscopic model to describe the mirror-field interaction, from which the dielectric response of the mirror to radiation are obtained from first principles. By using second-order perturbation theory, we adopted an open quantum system strategy to derived the master equation that describes the mirror's dynamics.
At the microscopic level, the effects of the back-reaction are encoded in coloured noise and non-local dissipation, which are accounted for by the corresponding fluctuation and dissipation kernels that enter in the definition of the time-dependent coefficients of the master equation for the mirror. At a macroscopic level, the back-reaction manifests by inducing mechanical friction and diffusion. We demonstrated that noise and fluctuation acting on the mirror satisfy standard fluctuation-dissipation relations, which in turn result in a direct relation between the stationary values of the friction and diffusion coefficients. Interestingly, we demonstrated that the physical mechanism from which noise and dissipation originates is the emission of particles in the field, by dynamical Casimir effect.
This work represents the first of a series of papers aimed to investigate the interaction between quantum fields and mechanical motion, within an open quantum systems framework. The results and the theory developed in this work are preparatory to the investigation of more complex and realistic configurations involving multiple mechanical objects, with the objective of studying the mechanism by which the vacuum fluctuations of the electromagnetic field transfer motion and mechanical energy between physically separated objects, as well as investigate viable techniques for using vacuum mediated coherent coupling for manipulating and transferring quantum states of mechanical motion between mesoscopic objects.

\section{Acknowledgements}
Enlightening discussions with Stephen Barnett, Miles Blencowe, James Cresser and Bei-Lok Hu are warmly acknowledged. This research was supported by the Leverhulme Trust Grant No. ECF-2019-461, and by University of Glasgow via the Lord Kelvin/Adam Smith (LKAS) Leadership Fellowship.

\bibliography{TwoMirrors.bib}

\end{document}